\documentclass[twocolumn,aps,showpacs,superscriptaddress]{revtex4}
\usepackage{amsmath}
\usepackage{amssymb}
\usepackage{epsf}
\usepackage{graphicx}

\begin{document}

\title{Characterization of graphene through anisotropy of constant-energy
maps in angle-resolved photoemission}

\author{\firstname{M.} \surname{Mucha-Kruczy\'{n}ski}}
\affiliation{Department of Physics, Lancaster University,
Lancaster, LA1 4YB, UK}

\author{\firstname{O.} \surname{Tsyplyatyev}}
\thanks{Present address: Department of Physics, University of Basel, Klingelbergstrasse 82,
4056 Basel, Switzerland} \affiliation{Department of Physics,
Lancaster University, Lancaster, LA1 4YB, UK}

\author{\firstname{A.} \surname{Grishin}}
\affiliation{Department of Physics, Lancaster University,
Lancaster, LA1 4YB, UK}

\author{\firstname{E.} \surname{McCann}}
\affiliation{Department of Physics, Lancaster University,
Lancaster, LA1 4YB, UK}

\author{\firstname{Vladimir I.} \surname{Fal'ko}}
\affiliation{Department of Physics, Lancaster University,
Lancaster, LA1 4YB, UK}
\affiliation{LPS-CNRS, University of Orsay,
Orsay, France}

\author{\firstname{Aaron} \surname{Bostwick}}
\affiliation{Advanced Light Source, MS 6-2100, Lawrence Berkeley
National Laboratory, Berkeley, CA 94720}

\author{\firstname{Eli} \surname{Rotenberg}}
\affiliation{Advanced Light Source, MS 6-2100, Lawrence Berkeley
National Laboratory, Berkeley, CA 94720}

\begin{abstract}
We show theoretically how constant-energy maps of the
angle-resolved photoemission intensity can be used to test wave
function symmetry in graphene. For monolayer graphene, we
demonstrate that the observed anisotropy of ARPES spectra is a
manifestation of what has been recently branded as electronic
chirality. For bilayer graphene, we show that the anisotropy of
the constant-energy maps may be used to extract information about
the magnitude and sign of interlayer coupling parameters and about
symmetry breaking inflicted on a bilayer by the underlying
substrate.
\end{abstract}

 \maketitle


\section{Introduction}


A similarity between electrons in monolayer graphene and
relativistic massless particles has been broadly discussed in the
literature
\cite{wallace,slonweiss,mcclure,DiVincenzo,Semenoff,AndoNoBS}. The
combination of a sublattice composition of electronic Bloch states
(treated as an ``isospin'') in a single atomic sheet of graphite
with a linear dispersion in the vicinity of corners of the
Brillouin zone makes them chiral, like Dirac fermions.
Experimentally, the chiral nature of charge carriers has been
deduced from a peculiar sequencing of plateaus in the quantum Hall
effect \cite{novo04,novo05,zhang05,novo06}, while the linear
dispersion relation has been observed directly by angle-resolved
photoelectron spectroscopy (ARPES)
\cite{bost07,zhou06,ohta06,ohta07,bost07b,zhou07,bost07c,bost07d}.
ARPES has already been used to provide information about the form
of the dispersion curves, renormalization of spectra by
electron-electron and electron-phonon interactions, and
information about quasiparticle lifetimes in the material
\cite{bost07,park07,cal07,tse07,pol07}. Here, we point out that
constant-energy angular maps of photoemission reflect the
chirality of electrons in graphene. For monolayers, we show that
the recently published ARPES data provides evidence for the
chirality of carriers in this material. We demonstrate that the
anisotropy of the constant-energy maps may be used to extract
information about the magnitude and sign of interlayer coupling
parameters in bilayer graphene and about the types of
symmetry-breaking effects produced by the underlying substrate or
doping. In particular, we demonstrate that one can distinguish
between two effects that may generate a gap in the bilayer
spectrum: interlayer asymmetry
\cite{mcc06,ohta06,guinea06,mcc06b,min07,castro,aoki07,mcc07,guinea07,mcc07b,guinea07b}
and symmetry breaking in the bottom layer resting on a SiC
substrate \cite{bost07b,zhou07,varchon07}.

Whereas the chirality of a relativistic particle is defined by its
spin, chirality in graphene refers to the sublattice composition
of plane-wave states of Bloch electrons. The honeycomb lattice of
monolayer graphene, Fig.~1(a), has two sites in the unit cell,
labeled as $A$ and $B$, and there is a degeneracy point at each of
two inequivalent corners $\mathbf{K}_{\pm} = \pm (4\pi /3a,0)$ of
the hexagonal Brillouin zone, also referred to as valleys,
Fig.~1(b). Near the center of valley $K_{+}$ in monolayer
graphene, electrons are described by a Dirac-like Hamiltonian,
\begin{eqnarray}
\hat{\mathcal{H}}_1 \approx \hbar v \mathbf{\sigma}.\mathbf{q} ,
\label{hc}
\end{eqnarray}
which determines the linear dispersion $\epsilon= \hbar v q$ of
electrons in the conduction band and $\epsilon=- \hbar v q$ in the
valence band \cite{wallace}. The eigenstates $\Psi$, within a
single valley, have different amplitudes on adjacent $A$ and $B$
sites, and, following the example of relativistic physics, they
may be written as a two-component ``spinor'' $\Psi = ( \psi_A ,
\psi_B )$. The chirality of a relativistic particle is
right-handed if its spin points in the same direction as its
momentum, left-handed if its spin points in the opposite
direction. By analogy, the relative phase $\phi$ between the wave
function on sublattice atoms indicates the `isospin' vector
$\mathbf{\sigma} = ( \cos \phi , \sin \phi , 0)$ of the chiral
state $\Psi = ( e^{-i\phi/2} , e^{i\phi/2} )$ of quasiparticles in
graphene. ARPES \cite{himpsel,shirley} is exactly the tool to
visualize this state through the angular dependence of the emitted
photoelectron flux.

The proposed analysis is based upon the standard theory of
angle-resolved photoemission \cite{himpsel,shirley}. In an ARPES
experiment, incident photons with energy $\hbar\omega$ produce
photoelectrons whose intensity $I$ is measured in a known
direction as a function of kinetic energy $E_{\mathbf{p}}\approx
\hbar^2 (|\mathbf{p}_{\parallel }|^{2}+p_{z}^{2})/2m$
\cite{himpsel,shirley}: $\hbar \omega = E_{\mathbf{p}} + A -
\epsilon_{q}$ where $A$ is the work function and $\epsilon_{q}$ is
the energy of Bloch electrons in graphene. Conservation of
momentum ensures that the component of the momentum parallel to
the graphene surface $\hbar \mathbf{p}_{\parallel} = \hbar ( p_x ,
p_y )$ is equal to the quasi-momentum $\hbar \mathbf{K}_{\pm} +
\hbar \mathbf{q}$ of Bloch electrons near valley $K_{\pm}$:
\begin{eqnarray}
\mathbf{p}_{\parallel} = \mathbf{K}_{\pm} + \mathbf{q} +
\mathbf{G} \, , \label{mtmcons}
\end{eqnarray}
where a reciprocal lattice vector $\mathbf{G}= m_{1}
\mathbf{b}_{1}+ m_{2} \mathbf{b}_{2}$ is written in terms of
primitive reciprocal lattice vectors $\mathbf{b}_{1} = (2\pi /a ,
2 \pi / \sqrt{3}a)$, $\mathbf{b}_2 = (2\pi /a , -2 \pi /
\sqrt{3}a)$, and integers $m_{1}$, $m_{2} $.

As graphene has two inequivalent atomic sites, the angular
dependence of the intensity may be accounted for by considering
two-source interference (\`{a} la Young's double slits). Outside
the sample, at a position $\mathbf{R}_0$ relative to the midpoint
of the two sources, electronic waves $e^{[ i
\mathbf{p}.(\mathbf{R}_0 + \mathbf{u}/2) - i \phi /2]}$ and $e^{[
i \mathbf{p}.(\mathbf{R}_0 - \mathbf{u}/2) + i \phi /2]}$ from
adjacent $A$ and $B$ sites combine. This yields the intensity $I$
of the two-source interference pattern,
\begin{eqnarray}
I \sim \cos^2 \left[ \frac{\mathbf{p}_{\parallel}.\mathbf{u}}{2} - \frac{\phi
}{2} \right] ,  \label{est1}
\end{eqnarray}
where $\mathbf{u} = (0,a/\sqrt{3})$ is the separation of the
adjacent sites, and near each corner of the Brillouin zone
$\mathbf{p}_{\parallel}.\mathbf{u} \approx 2 \pi (m_1 - m_2)/3$.
The first term in the argument of Eq.~(\ref{est1}) is a phase
difference due to the different path lengths of electron waves
emitted from two sublattices, while the second term, $-\phi /2$,
arises from the relative phase of the electronic Bloch states on
$A$ and $B$ sublattices determined by the quasiparticle chirality.

\begin{figure}[t]
\includegraphics[width=0.3\textwidth]{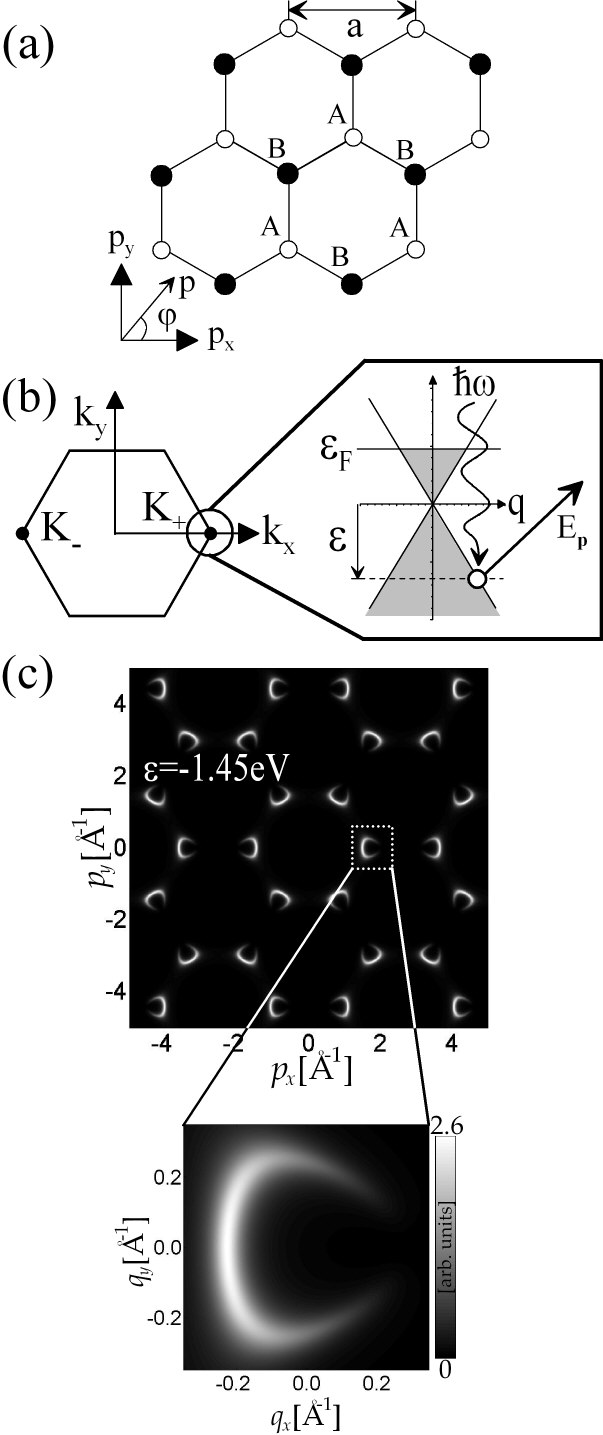}
\caption{(a) schematic of the monolayer lattice containing two
sites in the unit cell: $A$ (white circles) and $B$ (black). (b)
Schematic of the hexagonal Brillouin zone with two inequivalent
valleys $K_{\pm}$ and the low energy bands $\epsilon \approx \pm
\hbar v q$ near the $K_{+}$ point obtained by taking into account
intralayer hopping with velocity $v$. Shading indicates the region
of occupied states up to the Fermi energy $\epsilon_F$, dashed
line indicates a typical energy of states contributing to
photoemission, whereby incoming photons of energy $\hbar \omega$
produce photoelectrons of kinetic energy $E_{\mathbf{p}}$. (c) The
intensity of photoemission from states at a constant energy
$1.45$eV below the charge-neutrality point \cite{cnpoint} in
monolayer graphene, plotted as a function of photoelectron wave
vector $\mathbf{p}_{\parallel} = ( p_x , p_y )$ parallel to the
surface of graphene for $\mathbf{p}_{\parallel}$ covering several
Brillouin zones (top) and plotted as a function of photoelectron
wave vector $\mathbf{q} = ( q_x , q_y )$ in the vicinity of valley
$K_{+}$ (bottom) [note that the origin and scale of
$\mathbf{p}_{\parallel}$ and $\mathbf{q}$ are different]. Here we
use parameter values $\protect\gamma_0 = 3.0$eV, $s_0 = 0.129$,
$\Delta = 0$, and the energy width $\Gamma = 0.24$eV.}
\label{fig:1}
\end{figure}

\bigskip

\begin{tabular}{|c|c|c|c|}
  \hline
  energy & chirality & sublattice phase & ARPES \\
    $\epsilon_q$ & $\mathbf{\sigma}.\mathbf{n}_1$ & difference & anisotropy\\
  \hline
  $+vq$ & $+1$ & $\phi = \varphi$ & $I \sim \cos^2
(\varphi /2)$ \\
  $-vq$ & $-1$ & $\phi = \varphi + \pi$ & $I \sim \sin^2
(\varphi /2)$ \\
  \hline
\end{tabular}

\bigskip

{Table 1: Manifestation of electronic chirality in the anisotropy
of ARPES constant-energy maps in monolayer graphene in the valley
$K_{+}$, where angle $\varphi$ specifies the direction measured
from the center of the valley.}

\bigskip

Electrons in the conduction and valence bands at the valley
$K_{+}$, determined by the Dirac Hamiltonian, Eq.~(\ref{hc}),
differ by the projection of their isospin onto the direction of
their momentum $\mathbf{q} = (q\cos \varphi , q \sin \varphi)$ as
described by the chiral operator $\mathbf{\sigma}.\mathbf{n}_{1}$
where $\mathbf{n}_{1}(\mathbf{q})= (\cos \varphi , \sin \varphi)$:
$\mathbf{\sigma}.\mathbf{n}_1 = 1$ in the conduction band,
$\mathbf{\sigma}.\mathbf{n}_1 = - 1$ in the valence band, as
listed in Table~1. Note that, the first term in the argument of
Eq.~(\ref{est1}), arising from the path difference of electron
waves emitted from two sublattices, accounts for the relative
rotation in the interference pattern around the six corners of the
hexagonal Brillouin zone. Figure~1(c) shows a typical calculated
dependence of the intensity of the photoemission from states
(here, at energy $1.45$eV below the charge-neutrality point
\cite{cnpoint}) plotted as a function of wave vector
$\mathbf{p}_{\parallel }$, in agreement with the qualitative
prediction of the two-source interference picture
Eqs.~(\ref{hc},\ref{est1}), summarized in Table~1. The numerical
data, Fig.~1(c), appear to be consistent with experimentally
measured constant-energy maps \cite{bost07}.

So far, we have discussed the angular dependence of the
interference patterns, neglecting the effect of trigonal warping.
It leads to a triangular deformation of iso-energetic lines in the
band structure of graphene and $\epsilon_q (-\mathbf{q}) \neq
\epsilon_q (\mathbf{q})$ asymmetry of the electron dispersion
around each valley which becomes more pronounced for states
further from the charge-neutrality point. Another perturbation of
chiral particles in graphene may be asymmetry $\Delta
=\epsilon_{A} - \epsilon _{B}$ of on-site lattice energies
$\epsilon _{A}$, $\epsilon _{B}$\ due to the presence of a
substrate, leading to a gap $\Delta$ in the spectrum at low
energies. The presence of such asymmetry in graphene grown
epitaxially on SiC substrate, and the possibility of observing its
effect within spectroscopic accuracy, has recently been discussed
following experimental ARPES measurements of the low-energy band
structure \cite{bost07b,zhou07}. In Section~\ref{sect2} below, we
show that opening an $AB$ asymmetry gap in the monolayer spectrum
is accompanied by the loss of the chirality-related anisotropy of
ARPES angular maps at low energies, which can be used as an
additional test for the symmetry-breaking effect induced by a SiC
substrate.

In Section~\ref{sect3} we offer a detailed analysis of the
angle-dependent maps of ARPES of bilayer graphene, which is the
main goal of this work. First, we analyze angular photoemission
maps of an ideal ``pristine'' bilayer, taking into account
intricate details of its band structure using a tight-binding
model that employs the Slonczewski-Weiss-McClure parametrization
of relevant couplings \cite{slonweiss,mcclure}. In
Sections~\ref{sg1} and \ref{bichiral}, we show that angular maps
can be used to determine not only the magnitude, but also signs of
the interlayer coupling constants used in the tight-binding model.
If measured experimentally, the latter information may also prove
to be useful for general studies of bulk graphite. In
Section~\ref{a1} we analyze the influence of inter- and
intra-layer symmetry breaking in bilayers, and we show that the
effect of the interlayer charge transfer upon doping can be, in
principle, distinguished from crystalline asymmetry induced by a
SiC substrate.


\section{Photoemission from monolayer graphene}

\label{sect2}


To produce a quantitative prediction of the photoemission
intensity, we use Fermi's Golden Rule to calculate the probability
of a photo-stimulated transition from an initial band state with
2D quasi-momentum $\hbar\mathbf{k} = \hbar\mathbf{K}_{\pm } +
\hbar\mathbf{q}$ and energy $\epsilon_{\mathbf{q}}$ in graphene to
a continuum state with momentum $\hbar\mathbf{p}$ and energy
$E_{\mathbf{p}}$ in vacuum \cite{shirley}. The initial state wave
function in graphene is written as a linear combination of Bloch
wave functions on the $A$ and $B$ sublattices with coefficients
$\psi_{A}$ and $\psi_{B}$, respectively:
\begin{eqnarray*}
\Psi_{\mathbf{k}} \left( \mathbf{r} \right) = \sum_{j = A,B}
\psi_{j} \left( \mathbf{k} \right) \left[ \frac{1}{\sqrt{N}}
\sum_{\mathbf{R}_j} e^{i\mathbf{k}.\mathbf{R}_j} \Phi \left(
\mathbf{r} - \mathbf{R}_j \right) \right] ,
\end{eqnarray*}
where $\mathbf{R}_A$, $\mathbf{R}_B$ are the positions of $A$ and
$B$ type atoms, and $\Phi ( \mathbf{r})$ is a $p_z$ atomic
orbital. Then, the intensity $I$ of photoemission from states in a
given band may be written as
\begin{eqnarray}
I \sim |\Phi_{\mathbf{p}}|^{2} \left\vert \sum_{j} \psi_{j}
e^{-i\mathbf{G}.\mathbf{\tau }_{j}} \right\vert^{2}  \delta
\left(
E_{\mathbf{p}} + A - \epsilon_{q} - \omega \right) \, ,
\label{igen}
\end{eqnarray}
where $\Phi _{\mathbf{p}}=\int e^{-i\mathbf{p}.\mathbf{r}}\Phi (
\mathbf{r}) d^{3}r$ is the Fourier image of an atomic orbital
$\Phi \left( \mathbf{r}\right)$, and the wave vector component
parallel to the surface is conserved,
$\mathbf{q}=\mathbf{p}_{\parallel }- \mathbf{K}_{\pm} -
\mathbf{G}$, Eq.~(\ref{mtmcons}). The summation with respect to
index $j = \{ A , B \}$ takes into account coefficients $\psi
_{A}$ and $\psi _{B}$ located at atomic positions defined by basis
vectors $\mathbf{\tau}_A$ and $\mathbf{\tau}_B$ within a given
unit cell. The Dirac delta function, containing the work function
of graphene $A$, expresses energy conservation. Note that, in this
paper, we do not model dynamical effects that lead to energy
broadening \cite{bost07,park07,cal07,tse07,pol07}, but introduce a
Lorentzian $\delta ( \ldots ) \approx \Gamma / (\pi [ ( \dots )^2
+ \Gamma^2 ])$ in the figures with parameter $\Gamma$ representing
finite energy broadening.

A standard form \cite{saito,dressel02} of a tight-binding
monolayer Hamiltonian ${\hat{H}}_{1}$ and overlap-integral matrix
${\hat{S}}_{1}$ (that takes into account non-orthogonality of
orbitals on adjacent atomic sites), ${\hat{H}}_{1} \Psi =
\epsilon_{\mathbf{q}} {\hat{S}}_{1} \Psi$, is
\begin{eqnarray*}
{\hat{H}}_{1} &=& \left(
\begin{array}{cc}
\Delta /2 & - \gamma_{0} f \left( \mathbf{k}\right) \\
- \gamma_{0} f^{\ast} \left( \mathbf{k}\right) & -\Delta /2
\end{array}%
\right) , \\
{\hat{S}}_{1} &=& \left(
\begin{array}{cc}
1 & s_0 f\left( \mathbf{k}\right) \\
s_0 f^{\ast }\left( \mathbf{k}\right) & 1
\end{array}
\right) , \\
f(\mathbf{k}) &=& e^{i k_y a / \sqrt{3}} + 2 e^{-i k_y
a / 2 \sqrt{3}} \cos (k_x a/2) .
\end{eqnarray*}
Here, parameter $\gamma_0$ describes the strength of
nearest-neighbor hopping yielding the Fermi velocity $v=\left(
\sqrt{3}/2\right) a\gamma_{0}/\hbar$ \cite{gammas}, and $a$ is the
lattice constant. The parameter $s_0 \ll 1$ describes
non-orthogonality of orbitals, $\Delta =\epsilon_{A}-\epsilon
_{B}$ describes a possible asymmetry between $A $ and $B$ sites
(thus opening a gap $|\Delta |$). Note that here we neglected
next-nearest neighbor hops which do not produce any visible change
in the calculated spectra.

\begin{figure}[t]
\includegraphics[width=0.4\textwidth]{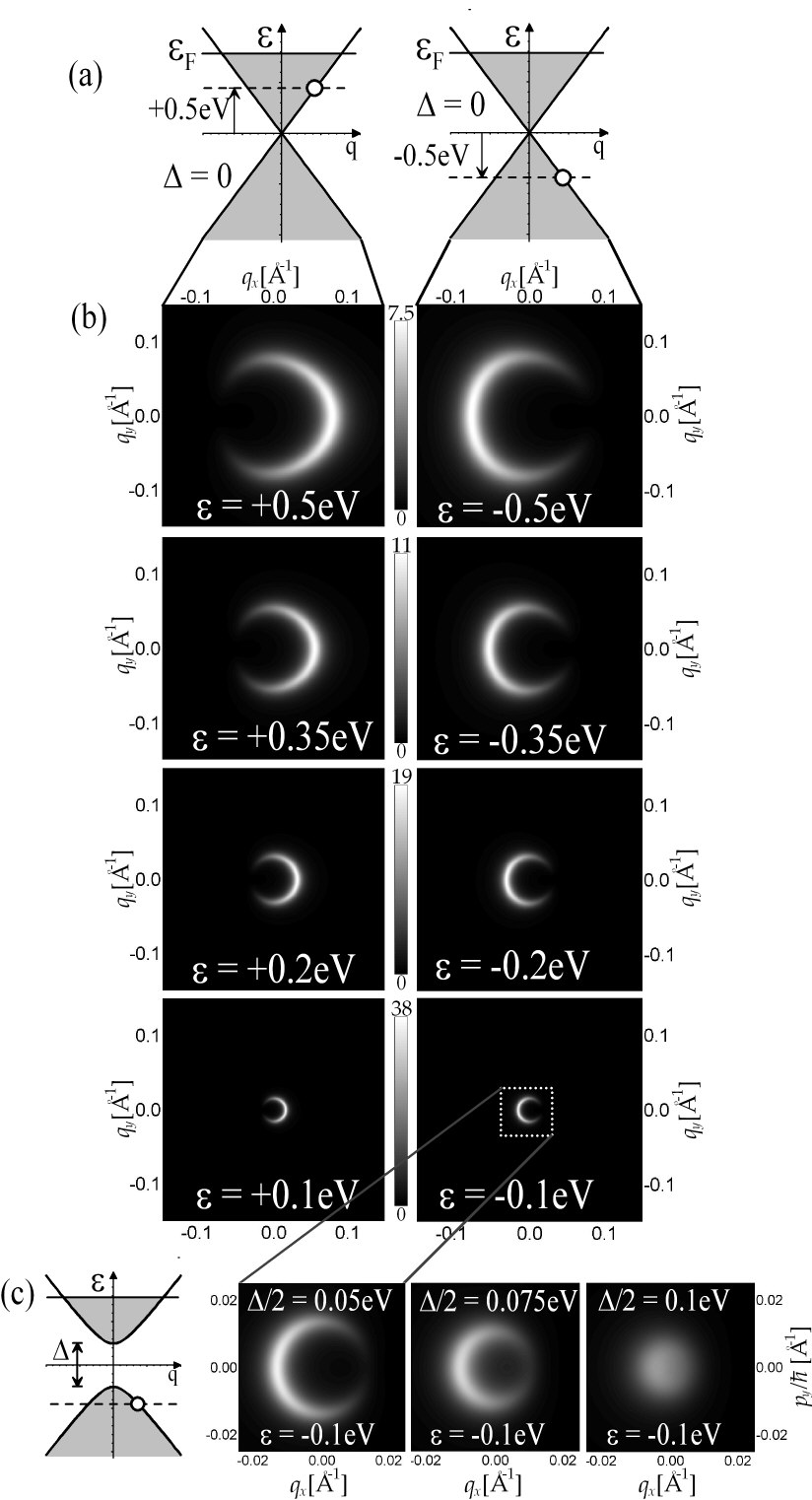}
\caption{(a) schematics of the low-energy bands  $\epsilon = \pm
\hbar v q$ near the $K_{+}$ point in the absence of intralayer
asymmetry $\Delta$. (b) The intensity of photoemission from states
at a fixed energy close to the charge-neutrality point in
monolayer graphene, plotted as a function of photoelectron wave
vector $\mathbf{q}_{\parallel} = ( q_x , q_y )$ parallel to the
surface of graphene in the vicinity of valley
$\mathbf{p}_{\parallel} = (4\pi /3a,0)$. Each plot corresponds to
a different energy with respect to the charge-neutrality point,
either above (left hand side) or below (right). Parameter values
are $\protect\gamma_0 = 3.0$eV, $s_0 = 0.129$, $\Delta = 0$, and
the energy width $\Gamma$ varies as the energy divided by six. (c)
The development of the intensity pattern for emission at a fixed
energy $0.1$eV below the charge-neutrality point in the vicinity
of valley $\mathbf{p}_{\parallel} = (4\pi /3a,0)$ in monolayer
graphene as intralayer asymmetry increases in magnitude $\Delta /2
= 0.05$, $0.075$, $0.1$eV. Parameter values are $\protect\gamma_0
= 3.0$eV, $s_0 = 0.129$, and the Lorentzian energy broadening
$\Gamma = 0.0167$.} \label{fig:2}
\end{figure}

Figure~1(c) shows constant-energy intensity patterns (``maps'') at
$1.45$eV below the charge-neutrality point \cite{cnpoint} in
monolayer graphene, plotted as a function of photoelectron wave
vector $\mathbf{p}_{\parallel} = ( p_x , p_y )$ parallel to the
surface of graphene, covering the whole Brillouin zone. Since the
patterns in the vicinity of each Brillouin zone corner are the
same, but rotated with respect to each other, we describe in
detail what is happening around one Brillouin zone corner.
Fig.~2(b) shows a series of plots demonstrating the evolution of
the constant-energy map with energy for the valley $\mathbf{K}_{+}
= (4 \pi / 3a , 0)$. Each plot is for a different fixed energy
from the charge-neutrality point with energies above (below) on
the left (right) hand side. For states above the charge-neutrality
point (left), the angular variation is $\cos^{2}(\varphi /2) $
where $\varphi$ is the angle of the momentum measured from the
center of the valley: comparison with Eq.~(\ref{est1}) yields
$\phi \equiv \varphi $, illustrating that the isospin is parallel
to the momentum $\mathbf{\sigma }. \mathbf{n}_{1}=1$. Fig.~2(b)
(right) shows the intensity for emission from states below the
charge-neutrality point in monolayer graphene. In this case, the
patterns are flipped with respect to those of the left and
comparison with Eq.~(\ref{est1}) yields $\phi =\varphi +\pi$
indicating that the isospin is antiparallel to the momentum
$\mathbf{\sigma }.\mathbf{n}_{1}=-1$.

Figure~2(c) shows the development of the fixed-energy intensity
pattern as asymmetry of on-site lattice energies $\Delta =
\epsilon_{A} - \epsilon _{B}$ increases, opening a small gap in
the spectrum. There are two principle effects on the ARPES
spectrum. For energy gaps $|\Delta |/ 2 \ll |\epsilon|$, the
mixing of the wave functions between $A$ and $B$ sites destroys
the perfect cancellation of the ARPES intensity, so that the ratio
between the maximum and minimum intensity becomes finite
[Fig.~2(c), left image]. As the gap increases towards the energy
probed, the ARPES contour becomes smaller and the intensity
anisotropy vanishes [Fig.~2(c), middle and right]. Thus, the
opening of an $AB$ asymmetry gap in the monolayer spectrum is
accompanied by the loss of the chirality-related anisotropy of
ARPES angular maps at low energies, which can be used not only as
a strong test for the symmetry-breaking effect induced by a SiC
substrate \cite{bost07b}, but also as a probe of wave function
mixing by $AB$ asymmetry or trigonal warping. Unlike $AB$
asymmetry, whose effects on the ARPES intensity are strongest near
the charge-neutrality point, trigonal warping affects the ARPES
spectral intensity only at very large energies.

For finite $\Delta$, an analytical, approximate description of the
ARPES intensity can be developed as follows. For electronic
energies much less than the $\pi$-band width ($q a \ll 1$),
\begin{eqnarray}
f(\mathbf{k}) \approx - \frac{\sqrt{3}a}{2} ( q_x - i q_y)
 + \frac{a^2}{8} ( q_x + i q_y)^2  , \label{fkexp}
\end{eqnarray}
and
\begin{eqnarray*}
{\hat{\mathcal{H}}}_{1} \approx \left(
\begin{array}{cc}
\Delta /2 & v{\pi }^{\dag } -\mu ({\pi })^{2} \\
v{\pi } -\mu ({\pi^{\dag }})^{2} & -\Delta /2
\end{array}
\right) \! ;  \,\, \pi = \hbar q_x + i \hbar q_y ,
\end{eqnarray*}
where $\mu = \gamma_0 a^2 / 8\hbar^2$ describes the strength of
trigonal warping (we assume that $\mu \hbar q \ll v$). This
determines the spectrum
\begin{eqnarray}
\epsilon_{q} \approx s\sqrt{\hbar^{2}v^{2}q^{2}-2\xi \mu
v\hbar^{3}q^{3}\cos 3\varphi +\mu^{2}\hbar^{4}q^{4}+ \frac{\Delta
^{2}}{4}}, \label{ene1}
\end{eqnarray}
where $s=1$ ($s=-1$) stand for the conduction (valence) band
index, and, using Eq.~(\ref{igen}), to the ARPES angular-dependent
intensity
\begin{eqnarray}
I\!\!&\sim& \!\!|\Phi _{\mathbf{p}}|^{2}\left\{ 1+\left\vert
\frac{\hbar vq}{\epsilon _{q}}\right\vert \left[ \cos \left(
2\theta \right) -\frac{\xi \mu \hbar q}{v}
\cos \left( 2\theta -3\xi \varphi \right) \right] \right\}  \notag \\
&& \qquad \times \, \delta \left( E_{\mathbf{p}}+A-\epsilon
_{q}-\hbar \omega \right) \, \delta_{\mathbf{q} ,
\mathbf{p}_{\parallel }- \mathbf{K}_{\pm } - \mathbf{G}} \, ,
\label{i1}
\end{eqnarray}
where $\theta = \frac{\xi \varphi }{2}-\frac{\pi }{3}\left(
m_{1}-m_{2}\right) +\frac{\pi }{4}\left( 1-s\xi \right)$.
Eqs.~(\ref{ene1},\ref{i1}) contain the full dependence on valley
$\xi = \pm 1$ and reciprocal lattice vector $(m_1 , m_2)$ indices
\cite{notw}.


\section{Photoemission from bilayer graphene}

\label{sect3}



\subsection{The use of ARPES to determine the sign of interlayer coupling parameter $\gamma_1$}\label{sg1}


Bilayer graphene \cite{novo06,mcc06,ohta06} consists of two
coupled hexagonal lattices with inequivalent sites $A1,B1$ and
$A2,B2$ in the first and second graphene sheets, respectively,
arranged according to Bernal ($A2$-$B1$) stacking \cite{mcc06} as
shown in Fig.~3(a). As in the monolayer, the Brillouin zone has
two inequivalent degeneracy points $K_{\pm}$ which determine two
valleys centered around zero energy in the electron spectrum. Near
the center of each valley the electron spectrum consists of four
branches, Fig.~3(b), with two branches describing states on
sublattices $A2$ and $B1$ that are split from zero energy by about
$\pm |\gamma_1|$, determined by the interlayer coupling
$\gamma_1$, whereas two low-energy branches are formed by states
based upon sublattices $A1$ and $B2$.

To model bilayer graphene we use a tight-binding Hamiltonian
matrix ${\hat{H}}_{2}$ and overlap-integral matrix ${\hat{S}}_{2}$
that operate in the space of coefficients $\psi ^{T}=(\psi
_{A1},\psi _{B2},\psi _{A2},\psi _{B1})$ at valley $K_{+}$
\cite{mcc06,guinea06,part06}:
\begin{eqnarray}
{\hat{H}}_{2} \!\!&=& \! \!\left(
\begin{array}{cccc}
\epsilon_{A1} & \gamma _{3}f^{\ast }\left( \mathbf{k}\right)
& \gamma_{4} f\left( \mathbf{k}\right) & -\gamma_{0} f\left( \mathbf{k}\right) \\
\gamma _{3}f\left( \mathbf{k}\right) & \epsilon_{B2} & -\gamma_{0}
f^{\ast}\left( \mathbf{k}\right)
& \gamma_{4} f^{\ast}\left( \mathbf{k}\right) \\
\gamma_{4} f^{\ast}\left( \mathbf{k}\right)
& -\gamma _{0}f\left( \mathbf{k}\right) & \epsilon_{A2} & \gamma _{1} \\
-\gamma_{0} f^{\ast}\left( \mathbf{k}\right) & \gamma_{4}
f\left(\mathbf{k}\right) & \gamma _{1} & \epsilon_{B1}
\end{array}%
\right) \! , \nonumber \\
\hat{{S}}_{2} \!\!&=& \! \!\left(
\begin{array}{cccc}
1 & 0 & 0 & s_{0}f\left( \mathbf{k}\right) \\
0 & 1 & s_{0}f^{\ast }\left( \mathbf{k}\right) & 0 \\
0 & s_{0}f\left( \mathbf{k}\right) & 1 & s_{1} \\
s_{0}f^{\ast }\left( \mathbf{k}\right) & 0 & s_{1} & 1%
\end{array}%
\right) \! . \label{hfour}
\end{eqnarray}%
We adopt the notation of the Slonczewski-Weiss-McClure model
\cite{slonweiss,mcclure} that is often used to describe bulk
graphite, in order to parameterize the couplings relevant to
bilayer graphene \cite{gammas}. Nearest-neighbor coupling within
each plane is parameterized by coupling $\gamma_{0}$ [$v=\left(
\sqrt{3}/2\right) a\gamma _{0}/\hbar $] and interlayer $A2$-$B1$
coupling is described by $\gamma _{1}$. The parameter $\gamma
_{3}$ describes direct $A1$-$B2$ interlayer coupling which leads
to an effective velocity $v_{3}=\left( \sqrt{3}/2\right) a\gamma
_{3}/\hbar $\ representing the magnitude of trigonal warping,
particularly relevant at low energy [we assume that $\hbar v q \gg
\gamma _{1}(v_{3}/v)$]. Parameter $\gamma_4$ describes $A1$-$A2$
and $B1$-$B2$ interlayer hopping. Using $\gamma_{4} = 0.044$eV
\cite{dressel02,parameters} we found no noticeable effect of
$\gamma_4$ on the ARPES plots, and, for simplicity, we use
$\gamma_4 = 0$ throughout. Other weaker tunneling processes
including next-nearest neighbor hopping are also neglected. The
parameter $s_1$ describes non-orthogonality terms arising from
overlaps between orbitals on different layers. Following numerical
analysis, we found it also has no noticeable effect on the ARPES
plots \cite{parameters}. In the following angular maps we use $s_1
= 0$. We note that some works on bilayer graphene use different
definitions of the tight-binding parameters (for example,
$\gamma_3$ is defined with an additional minus sign in
\cite{mcc06,mcc07,mcc07b}, but it has no effect on their
conclusions).

\begin{figure}[t]
\includegraphics[width=0.45\textwidth]{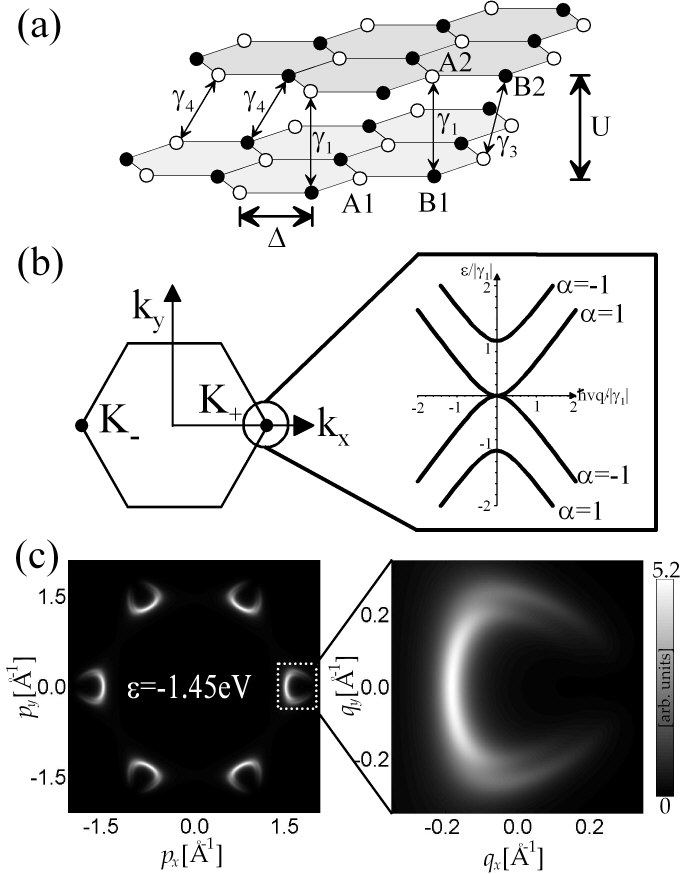}
\caption{(a) schematic of the bilayer lattice containing four
sites in the unit cell: $A1$ (white circles) and $B1$ (black) in
the bottom layer, and $A2$ (white) and $B2$ (black) in the top
layer.  (b) Schematic of the hexagonal Brillouin zone with two
inequivalent valleys $K_{\pm}$ and the low energy bands in the
absence of lattice asymmetry. The energy band index $\alpha = \pm
1$, Eq.~(\ref{alpeq}), is shown explicitly for the case $\gamma_1
< 0$. (c) The intensity of photoemission from states at a constant
energy $1.45$eV below the charge-neutrality point in bilayer
graphene, plotted as a function of photoelectron wave vector
$\mathbf{p}_{\parallel} = ( p_x , p_y )$ parallel to the surface
of graphene for $\mathbf{p}_{\parallel}$ covering the whole
Brillouin zone (left) and plotted as a function of photoelectron
wave vector $\mathbf{q} = ( q_x , q_y )$ in the vicinity of valley
$K_{+}$ (right) [note that the origin and scale of
$\mathbf{p}_{\parallel}$ and $\mathbf{q}$ are different]. Here we
use parameter values $\gamma_0 = 3.0$eV, $\gamma_1 = - 0.35$eV,
$\gamma_3 = - 0.15$eV, $\gamma_4 = 0.0$eV, $s_0 = 0.129$, $\Delta
= U = 0$, and the energy width $\Gamma = 0.24$eV.} \label{fig:3}
\end{figure}

The Bloch function amplitudes $\psi ^{T}=(\psi _{A1},\psi
_{B2},\psi _{A2},\psi _{B1})$ and band energy $\epsilon
_{\mathbf{q}}$, found using the Hamiltonian Eq.~(\ref{hfour}), can
be used to model the photoemission intensity. Figure~3(c) shows
constant-energy maps at $1.45$eV below the charge-neutrality point
\cite{cnpoint} in bilayer graphene, with the plot on the left hand
side showing values of $\mathbf{p}_{\parallel}$ covering the whole
Brillouin zone. The patterns in each valley are the same, but
rotated with respect to the others, so that we can focus on one of
them, highlighted in Fig~3(c). The anisotropy of the bilayer
pattern at this energy is similar to that of the monolayer,
Fig.~1(c), because the energetic width $\Gamma \sim |\gamma_1 |$
obscures features associated with the presence of two bands. To
observe differences between the two materials, we need to consider
the ARPES patterns at energies closer to the charge-neutrality
point.

\begin{figure}[t]
\includegraphics[width=0.45\textwidth]{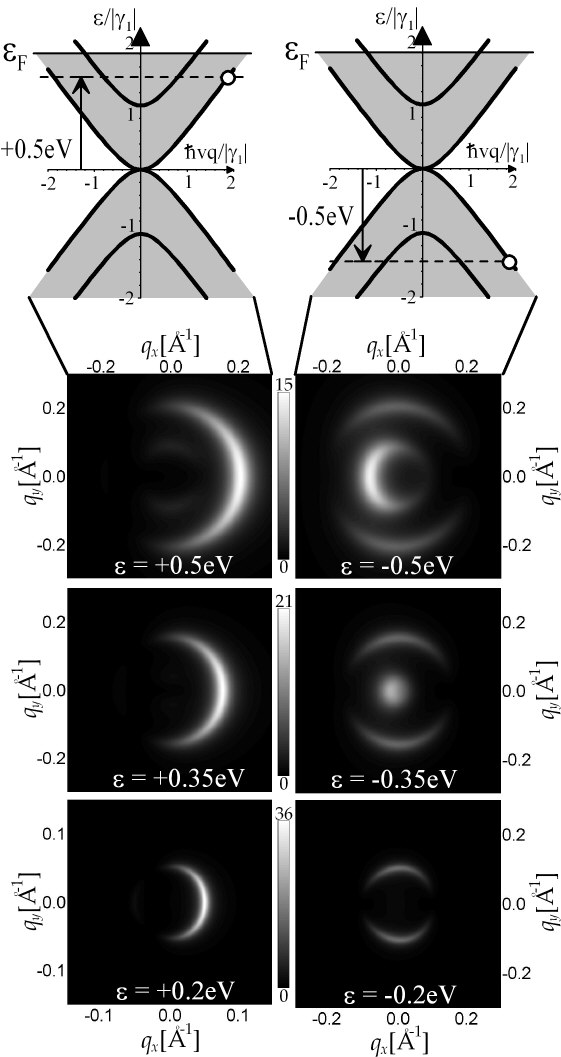}
\caption{The intensity of photoemission from states at a fixed
energy close to the charge-neutrality point in bilayer graphene,
plotted as a function of photoelectron wave vector $\mathbf{q} = (
q_x , q_y )$ parallel to the surface of graphene in the vicinity
of valley $\mathbf{p}_{\parallel} = (4\pi /3a,0)$. Each plot
corresponds to a different energy with respect to the
charge-neutrality point, either above (left) or below (right).
Parameter values are $\gamma_0 = 3.0$eV, $\gamma_1 = -0.35$eV,
$\gamma_3 = -0.15$eV, $\gamma_4 = 0.0$eV, $s_0 = 0.129$, $\Delta =
U = 0$, and the energy width $\Gamma$ varies as the energy divided
by six.} \label{fig:4}
\end{figure}

Figure~4 illustrates the evolution of the intensity pattern with
energy. At energies greater than the interlayer coupling,
$|\epsilon_q| > |\gamma_1|$ (the top two patterns), there are two
ring-like patterns, each corresponding to photoemission from
states in two bands, whereas, for low-energies, $|\epsilon_q| <
|\gamma_1|$ (the bottom two patterns), there is a single ring
corresponding to emission from the degenerate band only. Although
these plots have been obtained using a complete bilayer
Hamiltonian Eq.~(\ref{hfour}), it is convenient to discuss salient
features of the results in Fig.~4 using an analytic formula,
obtained by performing a linear-in-momentum expansion of
$f(\mathbf{k})$, Eq.~(\ref{fkexp}), and neglecting trigonal
warping due to $A1$-$B2$ interlayer coupling ($\gamma_{3}=0$),
$A1$-$A2$ and $B1$-$B2$ interlayer coupling ($\gamma_{4}=0$), and
non-orthogonality of orbitals ($s_{1} = s_{0} = 0$). In this case,
the four bands in the bilayer spectrum are described by
\begin{equation}
\epsilon_{q} \approx s {\textstyle\frac{1}{2}} |\gamma_{1}| \left[
\sqrt{1 + 4 \hbar^2 v^{2}q^{2}/\gamma _{1}^{2}} + b \right] ,
\label{ebs}
\end{equation}
where the parameters
\begin{eqnarray*}
b=\pm 1 \, ; \qquad s=\pm 1 \, ,
\end{eqnarray*}
identify the four bands: $b=1$ for the split bands with energy
$|\epsilon _{q}|\geq |\gamma _{1}|$ and $b=-1$ for the low-energy
`degenerate' bands that touch at zero energy, while $s=1$ ($s=-1$)
indicates the conduction (valence) bands. Then, the contribution
of a given band is
\begin{eqnarray*}
I\! \sim \! \frac{|\Phi_{\mathbf{p}}|^{2} g ( \varphi )} {\left[ 1
+ (\epsilon_{q}/\hbar vq)^2 \right]}  \, \delta \left(
E_{\mathbf{p}}+A-\epsilon _{q}-\hbar\omega \right)
\delta_{\mathbf{q} , \mathbf{p}_{\parallel} - \mathbf{K}_{\pm } -
\mathbf{G}}  ,
\end{eqnarray*}
where
\begin{eqnarray}
g ( \varphi ) &=& {\textstyle\frac{1}{2}} \left\vert e^{-i \varphi
} + \alpha e^{i \varphi }+ \frac{\epsilon_{q}}{\hbar vq} \left(
\alpha + 1 \right)
\right\vert^{2} \, , \label{ihigh} \\
&=& 1 + \alpha \cos ( 2 \varphi ) + \delta_{\alpha , 1} \!\!
\left[ \frac{4 \epsilon_{q}}{\hbar vq} \cos (\varphi ) + 2
\left(\frac{\epsilon_{q}}{\hbar vq} \right)^2 \right] , \notag
\end{eqnarray}
and
\begin{eqnarray}
\alpha = sb\gamma_{1} / |\gamma_{1}| .  \label{alpeq}
\end{eqnarray}
As the value of $\alpha$, Eq.~(\ref{alpeq}), depends on the sign
of the tight-binding parameter $\gamma_1$, comparison of the
angular dependence of $g ( \varphi )$ with experimental data
provides a method to determine the sign of $\gamma_1$
\cite{gammas}. To demonstrate this, we make a comparison with our
numerical data, plotted in Fig.~4. In this illustration, we assume
that $\gamma_1 < 0$, which is a natural choice given the $z
\rightarrow -z$ asymmetry of the $p_z$ orbitals of carbon. It
shows how, for this choice of the sign of $\gamma_1$, the
anisotropy of photoemission angular-maps differ in the split bands
and degenerate bands at energies above $\epsilon > 0$ and below
$\epsilon < 0$ the charge neutrality point. Note that changing the
sign of $\gamma_1$ to positive would lead to an interchange of
plots illustrating the ARPES behavior at $\epsilon > 0$ and
$\epsilon < 0$.

The most pronounced feature of the ARPES angular maps, depicted
for $\gamma_1 < 0$ in Fig.~4, is that, for energies $\epsilon > 0$
(left side of Fig.~4), photoemission spectra are dominated by
states in the degenerate bands, $b = -1$, which are nicely
described by the intensity profile $I \propto \cos^2 \varphi /2$.
In contrast, for $\epsilon < 0$ (valence bands, right side of
Fig.~4), ARPES intensity from the degenerate band $b = -1$ is
weak, whereas the split band, at energies $\epsilon < - | \gamma_1
|$, produces a bright, dominant signal. If observed
experimentally, such a behavior of ARPES maps in the conduction
and valence bands would be indicative of a negative sign of the
interlayer coupling $\gamma_1$ \cite{gammas}. If the
experimentally-observed constant-energy maps were interchanged for
negative and positive energies, it would be evidence for $\gamma_1
> 0$. Although the sign of $\gamma_1$ has directly observable
consequences for the ARPES pattern, tight-binding parameters for
graphite published so far have assumed $\gamma_1 > 0$
(\cite{char91} and references therein).


\subsection{Electron chirality in the ARPES of bilayer graphene and
the use of trigonal warping to determine the interlayer coupling
parameter $\gamma_3$}

\label{bichiral}

The behavior of low-energy particles in bilayer graphene is
perhaps even more remarkable \cite{novo06,mcc06,ohta06} than in a
monolayer. The low-energy bands (at energy $|\epsilon| \ll
|\gamma_1|$) have a parabolic energy-versus-momentum relation and
they support eigenstates of an operator
$\mathbf{\sigma}.\mathbf{n}_{2}$ with $\mathbf{\sigma}.
\mathbf{n}_{2} = 1$ for electrons in the conduction band and
$\mathbf{\sigma }.\mathbf{n}_{2} = -1$ for electrons in the
valence band, where $\mathbf{n}_{2}(\mathbf{q})= (\cos (2\varphi)
, \sin (2\varphi ))$, which means that they are chiral, but with a
degree of chirality different from that in the monolayer, with the
isospin linked to, but turning twice as quickly as, the direction
of momentum. An interpretation of the ARPES constant-energy maps
in terms of two-source interference Eq.~(\ref{est1}) predicts an
angular variation like $\cos^{2}(\varphi)$ for states above the
charge-neutrality point and $\cos^{2}(\varphi + \pi /2)$ for
states below (for $\gamma_1 < 0$).

\begin{figure}[t]
\includegraphics[width=0.45\textwidth]{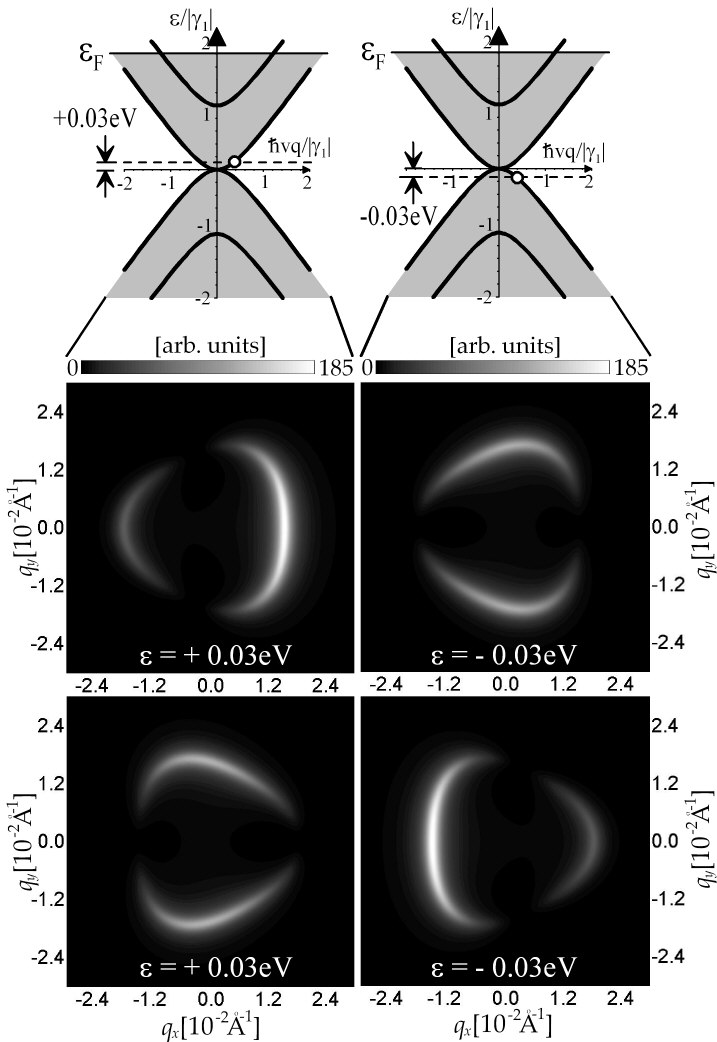}
\caption{The intensity of photoemission in bilayer graphene for
fixed energy very close to the charge-neutrality point in the
vicinity of valley $\mathbf{p}_{\parallel} = (4\pi /3a,0)$: for
states with energy $0.03$eV above the charge-neutrality point
(left) and states with energy $0.03$eV below the charge-neutrality
point (right). We consider different signs of the $A2$-$B1$
interlayer coupling strength, $\gamma_1$, with $\gamma_1 =
-0.35$eV (top) and $\gamma_1 = +0.35$eV (bottom). Other parameter
values are $\gamma_0 = 3.0$eV, $\gamma_3 = - 0.15$eV, $\gamma_4 =
0.0$eV, $\Gamma = 0.005$eV, and $s_0 = 0.129$ \cite{saito}.}
\label{fig:5}
\end{figure}

Figure~5 shows the calculated intensity of the photoemission in
the vicinity of valley $\mathbf{p}_{\parallel }=(4\pi /3a,0) $
from states very close to the charge-neutrality point in bilayer
graphene, at energy $0.03$eV above (left-hand side) and $0.03$eV
below (right). We consider two different signs of the $A2$-$B1$
interlayer coupling strength, $\gamma_1$, with $\gamma_1 =
-0.35$eV (top) and $\gamma_1 = +0.35$eV (bottom). For $\gamma_1 <
0$ and emission from the valence band (top right), the pattern is
like $\cos^{2}(\varphi + \pi /2)$ as expected for two-source
interference of chiral electron in bilayer graphene. As shown in
Eq.~(\ref{ihigh}) [and explained in detail in
Section~\ref{sect:discuss}], the intensity from this band is not
affected by corrections due to the presence of dimer $A2-B1$
orbitals (it has $\alpha = -1$). For emission from the conduction
band (top left side of Fig.~5), the interference pattern has two
peaks, but one of the peaks has about three times stronger maximum
intensity than the other because of the presence of the
contribution from dimer $A2-B1$ orbitals (this band has $\alpha =
+1$) \cite{discuss}. The bottom-left and bottom-right plots in
Fig.~5 show the constant-energy maps for $\gamma_1 >0$ for
emission above and below the charge-neutrality point,
respectively. In this case, the intensity pattern for emission
from the conduction band (bottom left) has two peaks with the same
maximum intensity, arising from the interference of waves from the
$A1$ and $B2$ sublattices. For emission from the valence band
(bottom right) the peaks have different maximum intensities, owing
to the interference of waves from four sublattices.

We note that, once the sign of $\gamma_1$ is known, the sign of
$A1$-$B2$ interlayer coupling $\gamma_3$ may also be deduced from
the orientation of trigonal warping of the intensity patterns near
the charge-neutrality point \cite{gammas}. In bilayer graphene,
there are two principle causes of trigonal warping. The first is
the presence of $A1$-$B2$ interlayer coupling $\gamma_3$ that will
tend to dominate at low energy, the second is higher-in-momentum
terms in the function $f(\mathbf{k})$ that will be important at
large energy. The latter causes trigonal warping in monolayer
graphene while the former is not present in a monolayer. At large
energies, when the higher-in-momentum terms dominate, the
orientation of trigonal warping is the same in bilayer graphene
[e.g. Fig.~3(c)] as in a monolayer [e.g. Fig.~1(c)] whereas, at
low energy, the orientation of trigonal warping in a bilayer
depends on the sign of parameter $\gamma_3$ (assuming that the
sign of $\gamma_1$ is known).

The orientation of trigonal warping flips on changing the sign of
$\gamma_1$ as seen by comparing the top and bottom plots in
Fig.~5. At very low energy, $\epsilon_{q} , \hbar v q \ll
\gamma_1$, and in the absence of lattice asymmetry, the energy
eigenvalues \cite{mcc06} are
\begin{eqnarray}
\epsilon_{q} \approx \pm \sqrt{ \hbar^2 v_3^2q^2 - 2 \xi
\frac{v_3v^2\hbar^3 q^3}{\gamma_1} \cos 3 \varphi + \frac{\hbar^4
v^4q^4}{\gamma_1^2}} , \label{tw}
\end{eqnarray}
where $v_3 = - \left( \sqrt{3}/2\right) a\gamma_{3}/\hbar $. This
expression illustrates that the angular dependent factor,
producing trigonal warping, depends on the sign of the ratio
$\gamma_3 / \gamma_1$. In this paper we usually choose $\gamma_1
<0$ and $\gamma_3 < 0$ to illustrate the possibility that the
orientation of trigonal warping is different at lower energies
[e.g. Fig.~5 (top)] than that at higher energies.


\subsection{Substrate-induced asymmetry in bilayer
graphene}\label{a1}


\begin{figure}[t]
\includegraphics[width=0.45\textwidth]{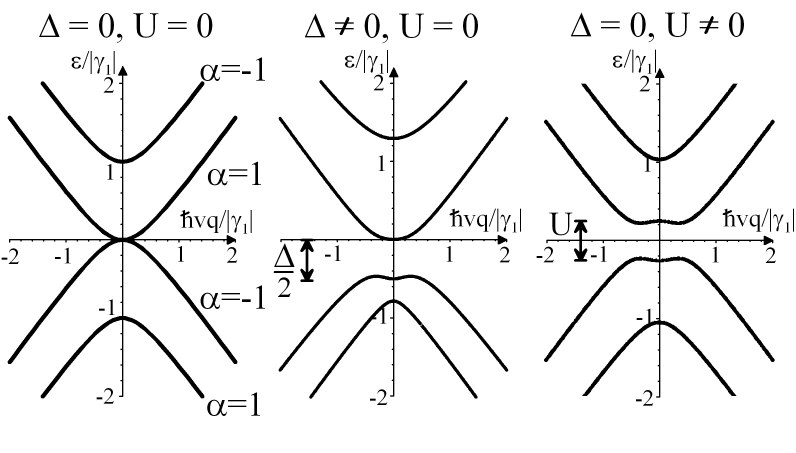}
\caption{The band structure of bilayer graphene in the vicinity of
a valley for no lattice asymmetry (left), substrate-induced
asymmetry $\Delta = \gamma_1 <0$ ($U=0$) (center), interlayer
asymmetry $U = \gamma_1 / 2$ ($\Delta=0$) (right). For clarity, we
use large values of asymmetry. The energy band index $\alpha = \pm
1$, Eq.~(\ref{alpeq}), is shown explicitly for the case $\gamma_1
< 0$ on the left hand side.} \label{fig:6}
\end{figure}

The Hamiltonian ${\hat{H}}_{2}$, Eq.~(\ref{hfour}), takes into
account the possibility of different on-site energies through its
diagonal components. Their effect may be understood by considering
the eigenenergies exactly at the center of the valley where
$f(\mathbf{k})=0$, namely $\epsilon = \epsilon_{A1}$, $\epsilon =
\epsilon_{B2}$, or
\begin{eqnarray*}
\epsilon = {\textstyle\frac{1}{2}}\left( \epsilon_{A2}+
\epsilon_{B1} \right) \pm \sqrt{{\textstyle\frac{1}{4}}
\left(\epsilon_{A2}- \epsilon_{B1}\right)^2 + \gamma_1^2} .
\end{eqnarray*}
Below, we distinguish between two types of asymmetry in bilayer
graphene parameterized using $\Delta =\epsilon _{A1}-\epsilon
_{B1}$ the difference between on-site energies of adjacent atoms
in the bottom layer due to the presence of a substrate, and
interlayer asymmetry $U=[(\epsilon_{A1} + \epsilon_{B1}) -
(\epsilon_{A2} + \epsilon_{B2})]/2$ between on-site energies in
the two layers arising from a doping effect and charge transfer to
the substrate
\cite{mcc06,ohta06,guinea06,mcc06b,min07,castro,aoki07,mcc07,guinea07,mcc07b,guinea07b}.

In Fig.~\ref{fig:6} (center), the band structure in the vicinity
of the K point is plotted  in the presence of substrate-induced
asymmetry $\Delta =\epsilon_{A1} - \epsilon_{B1}$ [the plot is
shown for $\gamma_1 <0$ and $\Delta <0$]. This type of asymmetry
introduces a gap $\sim |\Delta | /2$ as well as electron-hole
asymmetry. In Fig.~\ref{fig:6} (right), the band structure in the
vicinity of the K point is plotted in the presence of interlayer
asymmetry $U=[(\epsilon_{A1} + \epsilon_{B1}) - (\epsilon_{A2} +
\epsilon_{B2})]/2$. It does not break electron-hole symmetry, but
introduces a gap $\sim |U|$.

\begin{figure}[t]
\includegraphics[width=0.45\textwidth]{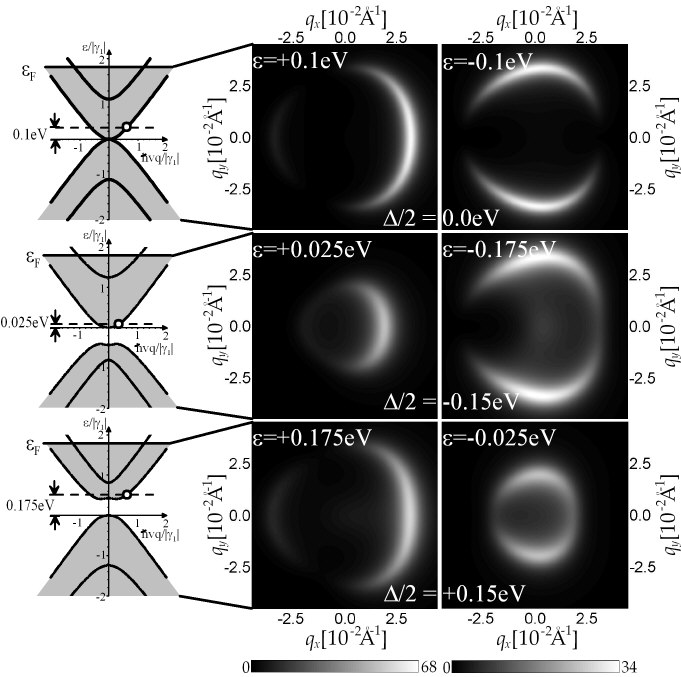}
\caption{Left (right) shows the development of the intensity
pattern for emission at a fixed energy $0.1$eV above (below) the
midgap energy in bilayer graphene in the vicinity of valley
$\mathbf{p}_{\parallel} = (4\pi /3a,0)$ as intralayer asymmetry
$|\Delta|$ increases [the energy with respect to the
charge-neutrality point is also indicated]. The plots show $\Delta
= 0$ (top), negative $\Delta/2 = -0.15$eV (middle), and positive
$\Delta/2 = 0.15$eV (bottom). Parameter values are $\gamma_0 =
3.0$eV, $\gamma_1 = - 0.35 $eV, $\gamma_3 = - 0.15$eV, $\gamma_4 =
0.0$eV, $s_0 = 0.129$, $\Gamma = 0.0167$eV.} \label{fig:7}
\end{figure}

As illustrated in  Fig.~7, the constant energy maps are sensitive
both to the magnitude and sign of the asymmetry $\Delta$. The
plots on the left (right) side of Fig.~7 show constant energy maps
for photoemission from conduction (valence) band states at energy
$0.1$eV above (below) the midgap energy. The top two plots are for
no asymmetry $\Delta = 0$, the middle two plots show negative
asymmetry $\Delta/2 = -0.15$eV, and the bottom two show positive
asymmetry $\Delta/2 = +0.15$eV. As for the monolayer, one effect
of asymmetry $\Delta$ is to impair the two-source interference
resulting in a weakening of the angular anisotropy of the
intensity pattern. The `Mexican hat' structure of the valence band
for negative $\Delta$, and the conduction band for positive
$\Delta$, is manifested in the larger ARPES contour for emission
from these states (shown on the middle right and the bottom left,
respectively) as opposed to their counterparts in the other band
(middle left and the bottom right, respectively). Experimentally,
such a difference in the size and nature of the ARPES contour for
emission from conduction or valence bands (at the same distance
from the midgap energy) would indicate the presence and sign of
intralayer asymmetry $\Delta$.

\begin{figure}[t]
\includegraphics[width=0.35\textwidth]{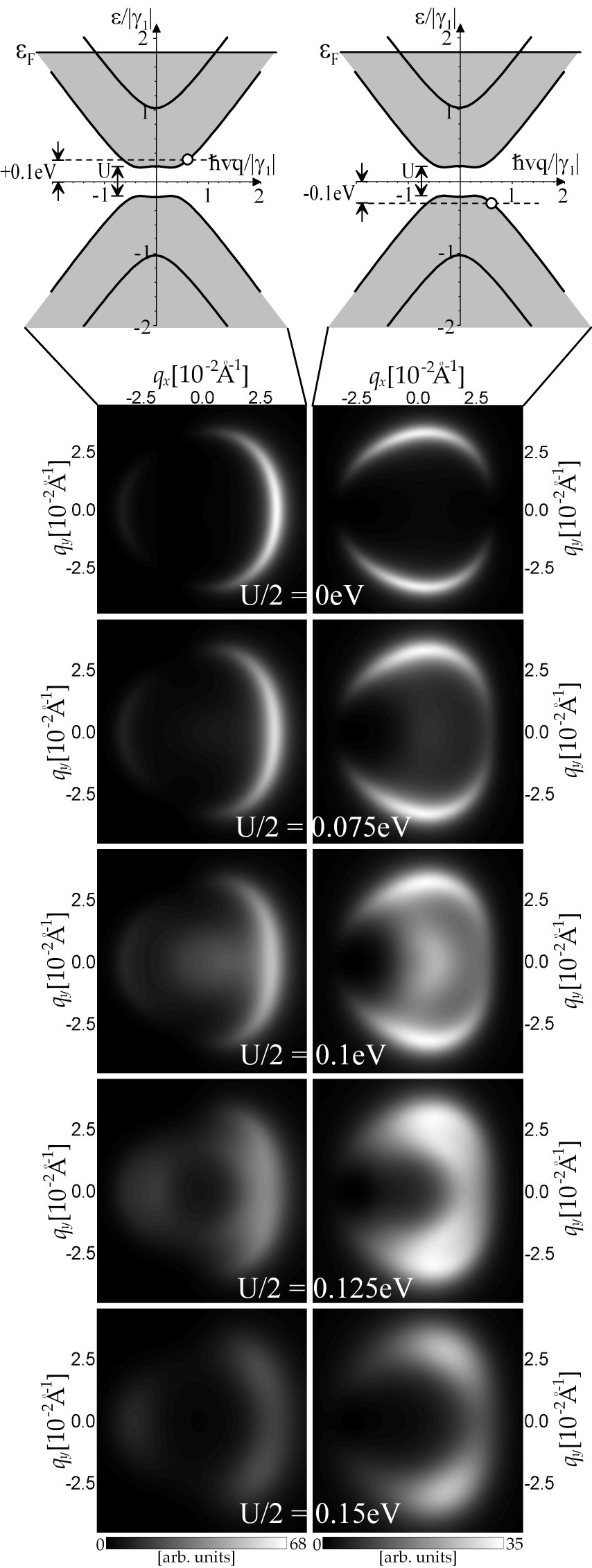}
\caption {The development of the intensity pattern in bilayer graphene in
the vicinity of valley $\mathbf{p}_{\parallel} = (4\pi /3a,0)$ as
interlayer asymmetry $U /2 = 0$, $0.075$, $0.1$, $0.125$, $0.15$eV
increases, for emission from states with energy $0.1$eV above the
charge-neutrality point on the left hand side, and energy $0.1$eV
below the charge-neutrality point on the right. Parameter values
are $\gamma_0 = 3.0$eV, $\gamma_1 = - 0.35 $eV, $\gamma_3 = -
0.15$eV, $\gamma_4 = 0.0$eV, $s_0 = 0.129$, $\Gamma = 0.0167$eV.}
\label{fig:8}
\end{figure}

Since interlayer asymmetry $U$ results in a gap $\sim |U|$ that
preserves electron-hole symmetry and does not depend on the sign
of $U$, the constant-energy photoemission maps are sensitive to
the magnitude of $U$ but not its sign. Figure~8 shows constant
energy maps for photoemission from conduction band states at
energy $0.1$eV above the charge-neutrality point (left hand side)
and from valence band states at energy $0.1$eV below the
charge-neutrality point (right) as $U$ increases in magnitude.
Generally, the effect of asymmetry $U$ is to weaken the angular
anisotropy of the intensity pattern. Both the conduction and
valence band in the vicinity of the charge neutrality point
display a `Mexican hat' structure \cite{mcc06,guinea06,mcc06b},
leading to relatively large ARPES contours (and additional
features as seen in the plot at $U/2 = 0.1$eV), in contrast to
asymmetry $\Delta$, Fig.~7, where the `Mexican hat' structure
appears in one band only.


\section{Discussion}\label{sect:discuss}


Our numerical data for bilayer graphene, Figs.~4,5, show that the
anisotropy of photoemission angular-maps differ in the split bands
and degenerate bands at energies above $\epsilon > 0$ (left side
of Fig.~4) and below $\epsilon < 0$ (right side of Fig.~4) the
charge neutrality point. These plots may be interpreted in terms
of the interference of photoelectron waves emitted from four
non-equivalent sublattices. In fact, for two of the four bands,
the parameter $\alpha = -1$ in Eq.~(\ref{ihigh}), so that the
contribution of orbitals on the `dimer' sites $A2$ and $B1$
cancel, leaving only the contribution of two terms $e^{\pm i
\varphi}$ in $g ( \varphi )$ that arise from orbitals on sites
$A1$ and $B2$. For the other two bands, $\alpha = 1$, the
contribution of orbitals on the `dimer' sites $A2$ and $B1$ to $g
( \varphi )$ do not cancel, but interfere with the contribution of
orbitals on sites $A1$ and $B2$, producing a different angular
dependence and greater peak intensity than for $\alpha = -1$.

As the value of $\alpha$, Eq.~(\ref{alpeq}), depends on the sign
of $\gamma_1$, comparison of the angular dependence of $g (
\varphi )$ with experimental data provides a method to determine
the sign of $\gamma_1$ \cite{gammas}. To demonstrate this, we make
a comparison with our numerical data, plotted in Fig.~4. For the
sign of $\gamma_1$ that we adopt in the numerics ($\gamma_1 <0$),
the split band above the charge-neutrality point has $\alpha = -1$
so the intensity from this band appears as a very faint ring (that
of smaller radius) in the plot at energy $\epsilon = + 0.5$eV in
Fig.~4. The degenerate band at this energy, however, has $\alpha =
1$ so the intensity from it appears as the ring of larger radius
with larger peak intensity. As the energy drops below
$|\gamma_1|$, left side of Fig.~4, the contribution of the split
band disappears to leave only the ring arising from the degenerate
band with $\alpha = 1$. The energy has to approach the
charge-neutrality point before the contribution of the dimer sites
$A2$ and $B1$, small in the parameter $\epsilon_{q} /\hbar vq
\approx \sqrt{\epsilon_{q} /\gamma_1}$, weakens to reveal an
anisotropy pattern characteristic of two source interference in
bilayer graphene, as explained in Section~\ref{bichiral}
\cite{discuss}.

The picture is quite different for energies below the
charge-neutrality point (right side of Fig.~4). In this case the
split band has $\alpha = 1$ so that the intensity from it appears
as the ring (of smaller radius) with larger peak intensity at
energy $\epsilon = - 0.5$eV in Fig.~4. The degenerate band has
$\alpha = -1$ so the intensity from it appears as the fainter ring
(that of larger radius) at energy $\epsilon = - 0.5$eV. As the
energy increases above $- |\gamma_1|$, right side of Fig.~4, the
contribution of the split band disappears to leave only the ring
arising from degenerate band with $\alpha = -1$. This is why the
intensity pattern $\sim \cos^2 ( \varphi )$ is much easier to
detect below the charge-neutrality point than above it. In fact,
whether it is easily visible above or below the charge-neutrality
point depends on the sign of $\gamma_1$ (here we chose $\gamma_1
<0$) so the experimental observation of the anisotropy $\sim
\cos^2 ( \varphi )$ will provide a way to determine the sign of
$A2$-$B1$ interlayer coupling $\gamma_1$ in bilayer graphene.

Finally, we note that the anisotropy of the constant-energy maps
may be influenced by other factors not modeled here. For example,
when the component of photoelectron momentum perpendicular to the
bilayer sample $p_z$ is large, we expect that photoelectron waves
emitted from the bottom layer will be attenuated with respect to
those emitted from the top layer. To obtain an impression of the
typical kind of effect, we introduced an exponential attenuation
[described by factor $\exp (-2z + 2i\beta)$ where $(z, \beta )$
are real parameters] of waves from the bottom layer. As shown in
Fig.~9 for photoemission from states at energy $0.1$eV below the
charge-neutrality point in bilayer graphene, the attenuation
results in a destruction of the double-peaked intensity pattern
and the phase factor $\beta$ has the effect of rotating the whole
pattern.

\begin{figure}[t]
\includegraphics[width=0.45\textwidth]{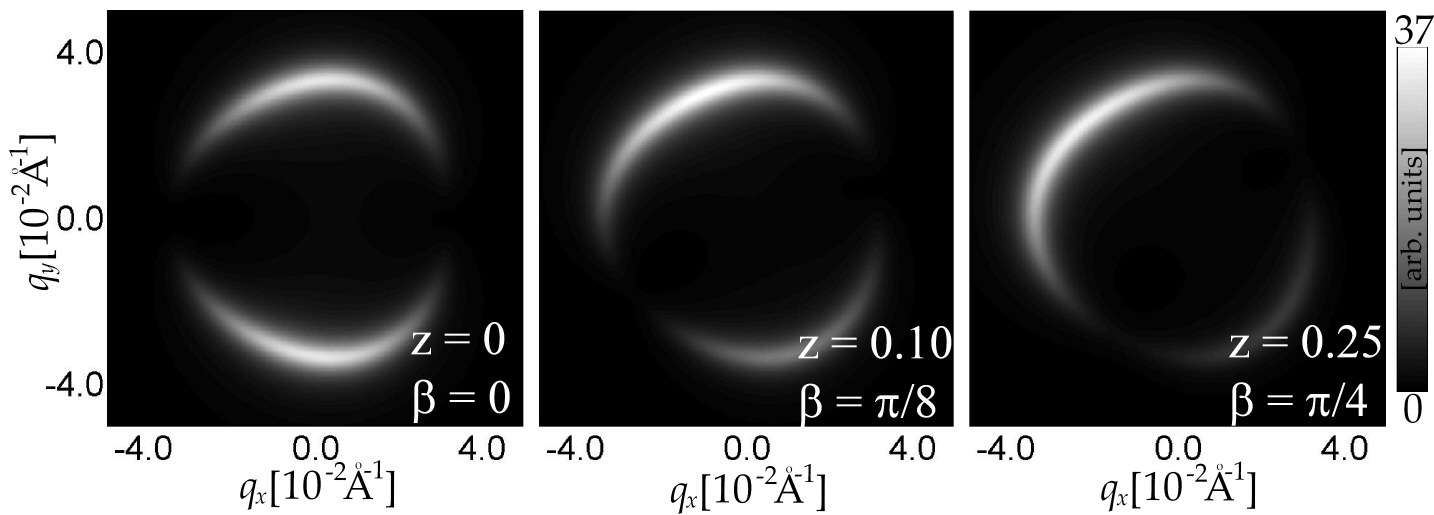}
\caption {The development of the intensity pattern in bilayer graphene for
emission from states with energy $0.1$eV below the
charge-neutrality point in the vicinity of valley
$\mathbf{p}_{\parallel} = (4\pi /3a,0)$ as attenuation [described
by factor $\exp (-2z + 2i\beta)$ where $(z, \beta )$ are real
parameters] of waves from the bottom layer, increases. Parameter
values are $\gamma_0 = 3.0$eV, $\gamma_1 = - 0.35 $eV, $\gamma_3 =
- 0.15$eV, $\gamma_4 = 0.0$eV, $s_0 = 0.129$, $\Gamma =
0.0167$eV.} \label{fig:9}
\end{figure}


\section{Conclusions}


Using Fermi's Golden Rule, we modeled the anisotropy of the
intensity of photoemission constant-energy maps at low energy in
graphene and demonstrated that the anisotropy is a manifestation
of electronic chirality. In monolayer graphene, photoemission may
be viewed as a two-source interference experiment, \`{a} la
Young's double slits, the sources being two inequivalent lattice
sites in the unit cell. The resulting intensity $\sim \cos^2 (
\varphi /2 )$ displays a single-peaked dependence on the direction
of momentum described by angle $\varphi$. In bilayer graphene, the
interference of emitted photoelectron waves from four atomic sites
produces single- or double-peaked constant-energy maps, depending
on the energy of the initial state in graphene. The marked
contrast between the anisotropy for emission from the conduction
or the valence band at energies below the $A2$-$B1$ interlayer
coupling strength, parameterized by $\gamma_1$ \cite{gammas},
provides an experimental method to determine the magnitude and
sign of parameter $\gamma_1$.

The shape of the photoemission constant-energy maps is determined
by the trigonal warping effect in graphene. In monolayers and
bilayers, the iso-energetic line changes from an almost-circular
to a triangularly-warped shape as the energy increases: the extent
of such warping is controlled by the dimensionless parameter $q
a$, where $q$ is the magnitude of the wave vector measured from
the center of the valley and $a$ is the lattice constant. In
bilayer graphene, strong trigonal warping may also occur at low
energy because of $A1$-$B2$ interlayer coupling, parameterized by
$\gamma_3$ \cite{gammas}, and the observation of this latter
trigonal warping provides an experimental method to determine the
magnitude and sign of parameter $\gamma_3$.

Measurement of the anisotropy of the intensity of photoemission
constant-energy maps provides a method to characterize realistic
graphene samples. As an example, we take into account
substrate-induced asymmetry that impairs the two-source
interference in monolayer graphene, resulting in a weakening of
the angular anisotropy of the intensity pattern. Analysis of
recent experimental data \cite{bost07b,zhou07} in terms of the
anisotropy of constant-energy maps may help to shed light on the
possible presence of asymmetry in graphene grown epitaxially on
SiC substrate. In bilayers, both substrate-induced asymmetry and
interlayer asymmetry alter the interference pattern: we describe
measurable differences between them. This illustrates the
potential of photoemission in the future characterization of
few-layer graphene samples.


\section{Acknowledgements}

The authors thank B.L.~Altshuler, T.~Ando, F.~Guinea, and
A.~Lanzara for discussions. This project has been funded by EPSRC
Portfolio Partnership EP/C511743/1, EPSRC First Grant
EP/E063519/1, ESF FoNE project SpiCo EP/D062918/1, the Royal
Society, and the Daiwa Anglo-Japanese Foundation. The Advanced
Light Source is supported by the Director, Office of Science,
Office of Basic Energy Sciences, of the U.S. Department of Energy
under Contract No. DE-AC02-05CH11231. V.~Fal'ko also acknowledges
support from the Alexander von Humboldt Foundation and hospitality
of the University of Hannover.


\begin{thebibliography}{99}
\bibitem{cnpoint} We define the charge-neutrality point as the
position of the Fermi level in nominally undoped graphene and we
set this position as that of zero energy. In monolayer graphene,
this is often called the Dirac point.
\bibitem{gammas} The tight-binding parameters are defined as
$\gamma_0 = - \langle \Phi_{A1} | H | \Phi_{B1} \rangle = -
\langle \Phi_{A2} | H | \Phi_{B2} \rangle$ where $\Phi_{A1}$ is an
atomic orbital located on site $A1$, $\gamma_1 = + \langle
\Phi_{A2} | H | \Phi_{B1} \rangle$, $\gamma_3 = + \langle
\Phi_{A1} | H | \Phi_{B2} \rangle$, and $\gamma_4 = + \langle
\Phi_{A1} | H | \Phi_{A2} \rangle = + \langle \Phi_{B1} | H |
\Phi_{B2} \rangle$. These definitions agree with those of the
Slonczewski-Weiss-McClure model of graphite
\cite{slonweiss,mcclure}. Note that the parameters describing
interlayer hopping ($\gamma_1$, $\gamma_3$, and $\gamma_4$) enter
the Hamiltonian of the Slonczewski-Weiss-McClure model with an
additional factor of two that takes into account the greater
number of neighboring graphene planes in bulk graphite as compared
to bilayer graphene.
\bibitem{parameters} We studied the influence of parameters
$\gamma_4$ and $s_1$ on the intensity pattern for emission at a
fixed energy $0.1$eV below the charge-neutrality point in bilayer
graphene (as in Figs.~\ref{fig:7}, \ref{fig:8}, \ref{fig:9}). We
found no noticeable effect for values $\gamma_4 \alt 0.1$eV and
$s_1 \alt 0.2$.
\bibitem{discuss}
For emission from the conduction band at very low energy (top left
side of Fig.~5), the interference pattern has two peaks [roughly
like $\cos^{2}(\varphi)$], but one of the peaks has about three
times stronger maximum intensity than the other. The reason is
that corrections due to the presence of dimer orbitals are small
(in parameter $\epsilon_q /\hbar vq$), but finite (this band has
$\alpha = +1$). Their influence may be estimated by considering
the function $g ( \varphi ) = 1 + \cos ( 2 \varphi ) + 4
(\epsilon_{q}/\hbar vq) \cos (\varphi ) + 2 (\epsilon_{q}/\hbar
vq)^2$, Eq.~(\ref{ihigh}). Comparing the maxima of the two peaks,
at angles $\varphi = 0$ and $\varphi = \pi$, gives $g(0) / g(\pi)
= [ (1 + \epsilon_{q}/\hbar vq) / (1 - \epsilon_{q}/\hbar vq)]^2$.
For the energy considered on the top left side of Fig.~5, where
$\epsilon_{q} \approx \hbar^2 v^2 q^2 / \gamma_1$, then
$\epsilon_{q}/\hbar vq \approx (\epsilon_{q} / \gamma_1)^{1/2}
\approx 0.29$. So, although this is `small', it yields $g(0) /
g(\pi) \approx 3.34$.
\end{thebibliography}
\end{document}